\documentclass[%
 reprint,
 amsmath,amssymb,
 aps,
]{revtex4-2}

\usepackage{graphicx}% Include figure files
\usepackage{dcolumn}% Align table columns on decimal point
\usepackage{bm}% bold math
\usepackage{tabularray}
\begin{document}

\preprint{APS/123-QED}

\title{GeV-level $\gamma$-ray and positron beams produced by collisions of ultra-intense ultra-short laser on high-energy electron beam}%

\author{Wanqing Su}
\affiliation{College of Physics, Henan Normal University, Xinxiang 453007, China}
\affiliation{Shanghai Advanced Research Institute, Chinese Academy of Sciences, Shanghai 201210, China}
\affiliation{Shanghai Institute of Applied Physics, Chinese Academy of Sciences, Shanghai 201800, China}
\author{Chunwang Ma}
 \altaffiliation[Corresponding authors: ]{machunwang@126.com for Chunwang Ma and caoxg@sari.ac.cn for Xiguang Cao, respectively.}
\affiliation{College of Physics, Henan Normal University, Xinxiang 453007, China}
\affiliation{Institute of Nuclear Science and Technology, Henan Academy of Sciences, Zhengzhou 450015, China}
\author{Xiguang Cao}
 \altaffiliation[Corresponding authors: ]{machunwang@126.com for Chunwang Ma and caoxg@sari.ac.cn for Xiguang Cao, respectively.}
\affiliation{Shanghai Advanced Research Institute, Chinese Academy of Sciences, Shanghai 201210, China}
\affiliation{Shanghai Institute of Applied Physics, Chinese Academy of Sciences, Shanghai 201800, China}
\affiliation{University of Chinese Academy of Sciences, Beijing 100049, China}
\author{Guoqiang Zhang}
\affiliation{Shanghai Advanced Research Institute, Chinese Academy of Sciences, Shanghai 201210, China}
\affiliation{Shanghai Institute of Applied Physics, Chinese Academy of Sciences, Shanghai 201800, China}
\affiliation{University of Chinese Academy of Sciences, Beijing 100049, China}
\author{Yuting Wang}
\affiliation{College of Physics, Henan Normal University, Xinxiang 453007, China}

% \author{Wan-Qing Su$^{1,2}$}
% \author{Chun-Wang Ma$^{1,3}$}\email[Corresponding author.]{Email:machunwang@126.com}
% \author{Xi-Guang Cao$^{2,4,5}$}\email[Corresponding author.]{Email:caoxg@sari.ac.cn}
% \author{Guo-Qiang Zhang$^{2,4,5}$}
% \author{Yu-Ting Wang$^{1}$}

% \affiliation{$^{1}$College of Physics, Henan Normal University, Xinxiang 453007, China\\
% $^{2}$Shanghai Advanced Research Institute, Chinese Academy of Sciences, \textit{Shanghai} 201210, China \\
% $^{3}$Institute of Nuclear Science and Technology, Henan Academy of Sciences, \textit{Zhengzhou} 450015, China \\
% $^{4}$Shanghai Institute of Applied Physics, Chinese Academy of Sciences, \textit{Shanghai} 201800, China \\
% $^{5}$University of Chinese Academy of Sciences, Beijing 100049, China
% }

\begin{abstract} 
Based on collisions between the 100 PW laser and 8 GeV superconducting linear accelerator constructing at the Shanghai hard X-ray free electron laser system (SHINE), the building of GeV-level $\gamma$-ray as well as positron beams are proposed according to particle-in-cell simulations. Key processes are considered involving the nonlinear inverse Compton scattering for $\gamma$-ray generation and the multiphoton Breit-Wheeler process for electron-positron pair production. Regardless of laser polarization, the simulations indicate that $\gamma$-ray beams achieve energy up to 8 GeV, brilliance around 10$^{27}$ photons/(s mm$^{2}$ mrad$^{2}$), and emittance as low as 0.1 mm mrad, while positron beams reach energy up to 7 GeV, brilliance around 4 $\times$ 10$^{24}$ positrons/(s mm$^{2}$ mrad$^{2}$), and emittance as low as 0.1 mm mrad. Various applications could benefit from the possible high-energy $\gamma$-ray and positron beams built at the SHINE facility, including fundamental physics of strong-field quantum electrodynamics theory validation, nuclear physics, radiopharmaceutical preparation, and imaging, etc.
\end{abstract}

\maketitle

\section{\label{sec:introduction}Introduction}

The Shanghai High Repetition Rate X-ray Free Electron Laser and Extreme Light Facility (SHINE) will offer a photon beam with energy spanning from 0.4 to 25 keV, leveraging its MHz-level high repetition rate and fs-level ultra-short pulses, to achieve exceptionally high average brightness and peak brightness \cite{chen2022beam, huang2023ming, liu2023status}. As a fourth-generation X-ray light source, the SHINE facility will provide cutting-edge experimental platforms for scientists across diverse fields worldwide. The ultra-intense laser induced high-energy particle accelerators, taking advantages of compactness, tunability and high brightness of laser system \cite{di2012extremely, nedorezov2021nuclear}, open new opportunities to multidisciplinary researches including superheavy nuclei synthesis \cite{thirolf2011laser, wanqing2024multi}, quantum-mechanical processes \cite{atlas2017evidence}, fundamental particles \cite{badelek2004photon}, nuclear structure and photonuclear physics \cite{gari1981photonuclear, tarbert2014neutron, wei2023spagins}. With its 100 PW laser source of Station of Extreme Light (SEL) \cite{hu2021numerical,shao2020broad,wang2019high}, which is upgraded and constructed based on the existing Shanghai Superintense Ultrafast Laser Facility (SULF) of 10 PW and 1 PW, and the 8 GeV electron beam of superconducting linear accelerator (SLA) \cite{zhu2017sclf,gong2021beam}, various particle beams can be generated by laser target shooting of the ultra-intense ultra-short laser (UIUSL) system including the high-energy $\gamma$-rays, positron sources, and even heavy-ion beams.

Between the collisions of the UIUSL and the high-energy electron beam \cite{matsukado2001nonlinear, luo2014nonlinear}, the $\gamma$-rays are generated through the nonlinear inverse Compton scattering (NICS), and the electron-positron pairs are yielded under the quantum electrodynamics (QED) effects through the multiphoton Breit-Wheeler process (MBWP) when the optical laser field is strong enough \cite{breit1934collision, burke1997positron, abramowicz2021conceptual}. Several experiments have successfully obtained $\gamma$-ray beams through NICS. The electron beam accelerated by a laser intensity of 4 $\times$ 10$^{19}$ W/cm$^{2}$ undergoes NICS with the laser intensity of 8 $\times$ 10$^{18}$ W/cm$^{2}$ in the Rutherford Appleton laboratory, resulting in the $\gamma$-ray beam with the maximum energy of 18 MeV and 1.8 $\times$ 10$^{20}$ photons/(s mm$^{2}$ mrad$^{2}$ 0.1$\%$ BW) peak brilliance \cite{sarri2014ultrahigh}. The electron beam accelerated by a laser intensity of 7.7 $\times$ 10$^{18}$ W/cm$^{2}$ also undergoes NICS with the laser intensity of 1.3 $\times$ 10$^{21}$ W/cm$^{2}$ in the Rutherford Appleton Laboratory, resulting in the $\gamma$-ray of critical energy 30 MeV \cite{cole2018experimental}. Earlier theory and simulations demonstrated that the increase of electron Lorentz factor and laser intensity could increase the energy of $\gamma$-ray produced, the enhanced stability of electron beam might reduce the emittance of the $\gamma$-ray beam \cite{mangles2006laser, valialshchikov2021narrow}, and the increase of laser intensity might also enhance the laser energy conversion efficiency to $\gamma$-rays and positrons \cite{zhao2022all}.

Combining all favorable factors above, at the SHINE facility, the high-quality $\gamma$-ray and positron beam could be generated by SEL 100 PW laser and SLA 8 GeV electron beam through NICS and MBWP process. Compared to the GeV-level $\gamma$-rays produced by bremsstrahlung \cite{albert2016applications}, the NICS method makes it possible to produce the first class high-quality $\gamma$-ray source above GeV-level for the SHINE facility, and significantly improve the energy of $\gamma$-ray source at the nearby Shanghai laser electron gamma source (SLEGS) \cite{wang2022}. Motivated by these promising opportunities, the particle-in-cell (PIC) program SMILEI is employed to simulate the whole processes between the UIUSL and high-energy electron beam. The feasibility of this scheme is elucidated, which also fulfills the gap between the $\gamma$-ray and positron beam researches. The article is organized as follows. In Section~\ref{sec:model}, the key physical mechanisms and parameter settings of simulations are highlighted. The simulation results, including the spatial distribution, energy spectrum, and spatial electric field distribution of particles, are presented in Section~\ref{sec:result}. The beam parameters and applications of particles are discussed in Section~\ref{sec:beam}, and the main findings are summarized in Section~\ref{sec:conclusion}.

\section{\label{sec:model}Models and simulations}

The SMILEI toolkit, which is a collaborative, open-source, user-friendly PIC code, has been applied into a wide range of physics studies from relativistic laser-plasma interaction to astrophysical plasmas \cite{derouillat2018SMILEI}. Within all simulated grids, the motion of particles in the electromagnetic field satisfy the Vlasov's equation and gradually form a self-consistent dynamical system \cite{perelomov1966ionization, perelomov1967ionization, ammosov1986tunnel}. The mechanism of high-energy $\gamma$-rays production in collisions between the UIUSL and the high-energy electron beam is the incoherent process of NICS. The dynamics of a single electron with charge $-e$ and mass $m$ could be determined by the Lorentz equation in an arbitrary external field, which is described by the covariant form of the Lorentz equation using the electromagnetic field tensor $F^{\mu \nu }$ in the form of
\begin{equation}
\frac{\mathrm{d}{p}^{\mu }}{\mathrm{d}\tau } = -\frac{e}{mc}{F}^{\mu \nu }{p}_{\nu }.
\end{equation}
The Lorentz invariant quantum parameter for the electron is
\begin{equation}
\chi = \left | \frac{{F}^{\mu \nu }}{{E}_{s}}\frac{{p}_{\nu }}{mc}\right | = \frac{\gamma_e}{{E}_{s}}\sqrt{{\left (\mathbf{E} + \mathbf{v}\times \mathbf{B}\right )}^{2}-\frac{{\left ( \mathbf{v}\cdotp \mathbf{E}\right )}^{2}}{{c}^{2}}},
\end{equation}
and the Lorentz invariant quantum parameter for the photon at the time of photon emission is denoted as
\begin{equation}
{\chi}_{\gamma } = \frac{{\gamma}_{\gamma } }{{E}_{s}}\sqrt{{\left (\mathbf{E} + \mathbf{c}\times \mathbf{B}\right )}^{2}-\frac{{\left ( \mathbf{c}\cdotp \mathbf{E}\right )}^{2}}{{c}^{2}}},
\end{equation}
where $\gamma_e = \epsilon_e /\left ( {m}_{e}{c}^{2}\right )$ and ${\gamma}_{\gamma } = {\epsilon}_{\gamma } /\left ( {m}_{e}{c}^{2}\right )$ are the normalized energies of the radiating particle and emitted photon, respectively; $\mathbf{v}$ and $\mathbf{c}$ are their respective velocities, $c$ is the speed of light in vacuum. $\mathbf{E}$ and $\mathbf{B}$ denote the electric field and magnetic field, respectively; and ${E}_{s} = {m}^{2}{c}^{3}/\left ({\hbar}e\right )\simeq 1.3\times {10}^{18}$ V/m is the Schwinger field.

The MBWP is the process of high-energy photons decaying into an electron-positron pair in the intense electromagnetic field \cite{lobet2017generation, di2019improved}. The strength of the QED effects for electron and positron depends on the photon quantum parameter
\begin{equation}
{\chi }_{\gamma }^{BW} = \frac{{\gamma}_{\gamma } }{{E}_{s}}\sqrt{{\left ( \mathbf{{E}_{\perp }}+\mathbf{c}\times \mathbf{B}\right )}^{2}},
\end{equation}
where $\mathbf{{E}_{\perp }}$ is the electric field orthogonal to the propagation direction of the photon.

\begin{figure}[htbp]
\includegraphics{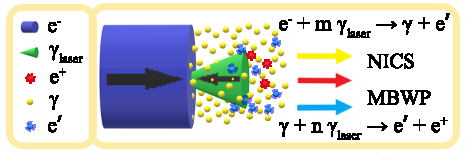}
\caption{(Color online) Schematic collision between the ultra-intense ultra-short laser ($\gamma _{laser}$) and the high-energy electron beam ($e^-$). The moving directions of particles are indicated by thick arrows.}
\label{theory}
\end{figure}

The parameters adopted in the SMILEI simulation are based on the SHINE facility, of which the monoenergetic electron beam density is 3.2 $\times$ 10$^{24}$ /m$^{3}$ with a beam spot size 10 $\times$ 10 $\mu $m$^{2}$. The laser is set to be a Gaussian laser of the wavelength $\lambda$ = 1 $\mu$m, with a focal spot size 5 $\mu$m, a full-width-at-half-maximum (FWHM) 15 fs, and a peak intensity $I={10}^{23}~\mathrm{W/{cm}^{2}}$, which corresponds to a normalized laser amplitude of ${a}_{cir}\approx$ 190 for circularly polarized laser (denoted by CPL) and ${a}_{lin}\approx$ 269 for linearly polarized laser (denoted by LPL). The spatial size of the two-dimensional and three-velocity (2D3V) particle-in-cell (PIC) simulation is set to be 300 $\times$ 100 $\mu$m$^{2}$, with a spatial step size of 100 nm and a time step of 22 as. Each grid initially contains 16 electrons in the electron beam region. The electron beam moves from the center on the left towards the right, while the laser focusing on the electron beam comes in from the center on the right towards the left. The $\gamma$-rays are emitted by electron and photons immediately at collision through the NICS mechanism \cite{timokhin2010time, lobet2016modeling}, specifically, the Monte Carlo simulation \cite{elkina2011qed, duclous2011monte, niel2018quantum}. The electron-positron pairs are created by photons through the MBWP mechanism almost simultaneously. The collision process, as well as the colliding equations are illustrated in Fig.~\ref{theory}, which includes
\begin{eqnarray}
     & \mathrm{NICS}: e^{-}+m\,\gamma_{laser}\rightarrow \gamma + e^{-}, \\
     & \mathrm{MBWP}: \gamma +n\,\gamma_{laser}\rightarrow e^{+}+e^{-}.
\end{eqnarray}

%figure1

\section{\label{sec:result}Results}

The results for particles productions, at the time when the electrons interact with the laser and the number of positrons no longer increases, are determined according to CPL and LPL patterns. The spatial distribution and energy of produced particles, and space electric field are presented.

\subsection{\label{sec:xy}Particle spatial distribution}

\begin{figure}[htbp]
\includegraphics{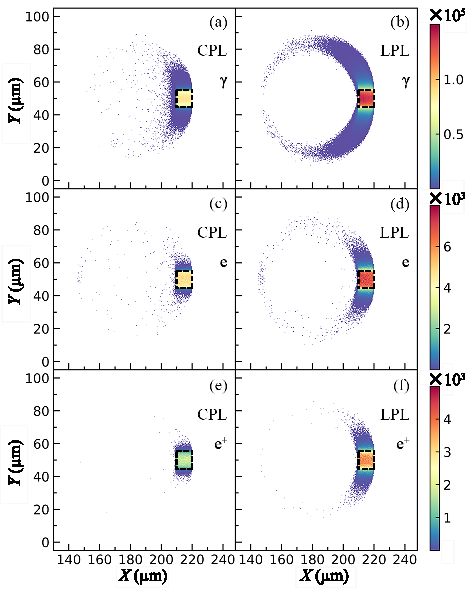}
\caption{(Color online) Particle spatial distributions at 700 fs of the SMILEI simulated collisions between 100 PW laser and 8 GeV electron beam based on the Station of Extreme Light (SEL) and superconducting linear accelerator (SLA) at the SHINE facility. (a) and (b) are for $\gamma$-rays production under circularly polarized laser (CPL) and linearly polarized laser (LPL) patterns, respectively. (c) and (d) are for electrons production under CPL and LPL, respectively. (e) and (f) are for positrons production under CPL and LPL, respectively. Particle beam parameters are given in Section~\ref{sec:beam}.}
\label{xy-pep}
\end{figure}

The spatial distributions for particles are plotted in Fig.~\ref{xy-pep}. For the CPL pattern, as shown in Fig.~\ref{xy-pep}(a), (c) and (e), the spatial distributions of $\gamma$-rays are more dispersed in the $Y$ direction compared to electrons and positrons. The spatial distributions of electrons and positrons almost overlap. The size of the beam center (black dashed box) almost equals to the initial electron beam size. The central density of photons is greater than that of electrons, and positrons have the lowest central density. For the LPL pattern, as shown in Fig.~\ref{xy-pep}(b), (d) and (f), the spatial distributions of particles resemble a crown adorned with a gem and the beam center as the ``gem on the crown'', which are quite different from the ``meteor'' pattern in the ``rotating forward'' CPL pattern. The $\gamma$-rays are also more dispersed in the $Y$ direction compared to electrons and positrons. $\gamma$-rays have the highest central density, followed by electrons, and then positrons similarly. Compared to the CPL pattern, the particles in the LPL pattern spread on sides of the beam center in vertical direction and have a higher particle density. The same phenomena happen in the $Y$ direction for the particles. Under both patterns, the particles in the horizontal direction are distributed mostly to the left of the beam center, with a cluster of electrons at the ``tail end'' which refers to the left end of the $X$ axis and the center of the $Y$ axis.

% figure2

\subsection{\label{sec:Ekin}Particle energy}

\begin{figure}[htbp]
\includegraphics{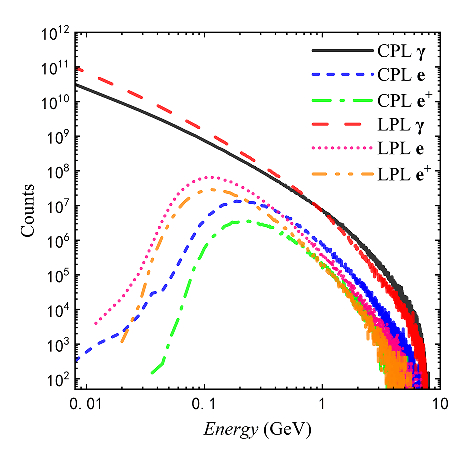}
\caption{(Color online) Particle energy spectrum at 700 fs of the SMILEI simulated collisions between 100 PW laser and 8 GeV electron beam based on the SEL and SLA at the SHINE facility. The CPL and LPL denote the circularly polarized laser and linearly polarized laser, respectively.}
\label{Ekin-pep}
\end{figure}

The energy spectra of produced particles are plotted in Fig.~\ref{Ekin-pep}. The energy spectrum of $\gamma$-rays decreases quickly in the range of $Energy\in$ [0, 8] GeV. For the LPL pattern, more photons are produced in the lower energy range ($Energy < $ 0.9 GeV), while for the CPL pattern, more photons are produced in the higher energy range ($Energy \geqslant $ 0.9 GeV), with the maximum energies for both patterns approaching 8 GeV. The electron energy broadens from the mono-energy to the low-energy region, and the energy spectra of electrons and positrons both gradually increase to form peaks and then decrease with the increasing energy. For the CPL pattern, the peaks of the electrons and positrons distributions both are located at $Energy = $ 0.2 GeV, and the peaks of the electrons and positrons both are located at $Energy =  $ 0.1 GeV for the LPL pattern, and more positrons and electrons are produced compared to the LPL pattern the peak positions. More electrons are produced in the LPL pattern than the CPL one is when $Energy < $ 0.5 GeV, while opposite trends happens when $Energy \geqslant $ 0.5 GeV. The number of positrons produced in the CPL pattern is similarly less than that of the LPL pattern when $Energy <$ 1.1 GeV, while the trend revers when $Energy \geqslant $ 1.1 GeV. Notably, the maximum energy of electrons still approaches 8 GeV, and the maximum energy of positrons is close to 7 GeV.

% figure3

\subsection{\label{sec:Ex}Space electric field}

\begin{figure}[htbp]
\includegraphics{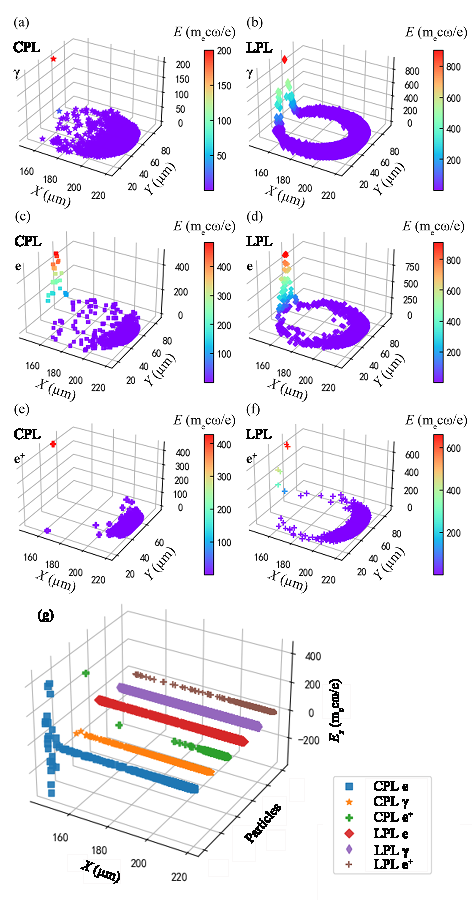}
\caption{(Color online) The spatial electric field distribution at 700 fs of the SMILEI simulated collisions between 100 PW laser and 8 GeV electron beam based on the SEL and SLA at the SHINE facility. The panels (a), (c), and (e) denote the $\gamma$-rays, electrons and positrons in the CPL pattern, respectively. The panels (b), (d), and (f) denote $\gamma$-rays, electrons and positrons in the LPL pattern, respectively. Panel (g) is for the electric field component distribution in the $Z$ direction.}
\label{E-pep}
\end{figure}

The electric field simulated by SMILEI for collisions between 100 PW laser and 8 GeV electron beam based on the SHINE facility is plotted in Fig.~\ref{E-pep}. The electric field strength reflects the spatial and momentum evolution of charged particles at their positions, offering a more detailed reference for intercepting particle beams. Charged particles exhibit greater stability in regions where the electric field remains stable. As shown in Fig.~\ref{E-pep}(a)-(f), the electric field reaches its maximum at the ``tail end'' of the particles, decreasing towards both sides of the $Y$ axis and the right side of the $X$ axis. For electrons and positrons, the electric field at the beam center is relatively low, enabling most particles to move forward collectively. Conversely, the electric field is higher at the``tail end'', causing particles at the rear to diverge outward. The electric field is apparently weaker in the CPL pattern compared to the LPL pattern.

Seen from Fig.~\ref{E-pep}(g), the electrons and positrons in the CPL pattern are subjected to an electric field force in the $Z$ direction, while in the LPL pattern, they are not subjected to any electric field force in the $Z$ direction. It is revealed that electrons and positrons at the ``tail end'' diverge in three dimensions in the CPL pattern, while they only diverge in the $X$ and $Y$ directions in the LPL pattern.

% figure4

\section{\label{sec:beam}Discussion}

To specifically discuss the characteristics of particle beams produced in the simulations, an area of $X \in$ [190, 230]~$\mu$m and $Y \in$ [45, 55]~$\mu$m is delimited for further  estimation of the angular spectrum, particle beams parameters, experimental layout and potential applications in scientific researches.

\subsection{\label{sec:beam-ang}Angular spectrum}

\begin{figure}[htbp]
\includegraphics{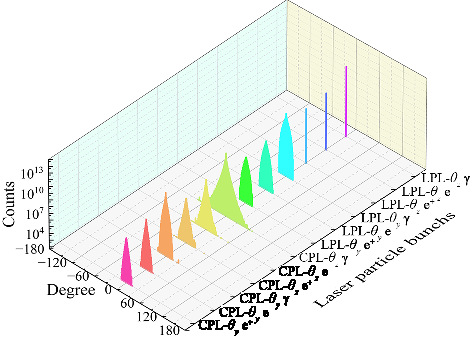}
\caption{(Color online) The angular spectrum of particle momentum (${\theta }_{y}=\arctan (p_y/p_x)$ and ${\theta }_{z}=\arctan (p_z/p_x)$) in the beam region at 700 fs simulated by SMILEI for collision between circularly polarized (CPL) and linearly polarized (LPL) SEL 100 PW laser and SLA 8 GeV electron beam.}
\label{A-beam-pep}
\end{figure}

The angular spectrum of particles in the selected beam region are plotted in Fig.~\ref{A-beam-pep}. For convenience, ${\theta }_{y}=\arctan (p_y/p_x)$ and ${\theta }_{z}=\arctan (p_z/p_x)$ are defined. One sees that the ${\theta }_{y}$ angular spectra for different particles in the CPL pattern are obviously narrower than those in the LPL pattern, with almost equal peak values (${\theta }_{y}\approx$ 0). Among different polarization patterns of lasers, the ${\theta }_{y}$ angular spectrum of $\gamma$-rays shows the widest distribution with the highest peak values (about $2.4\times 10^{10}$ for CPL and $7.7\times 10^{10}$ for LPL). The electrons show an narrower angular spectrum with lower peak values around $1.2\times 10^{9}$ for CPL and $6.8\times 10^{8}$ for LPL. Positrons display the narrowest ${\theta }_{y}$ angular spectrum with the lowest peak value (approximately $2.0\times 10^{8}$ for CPL and $2.6\times 10^{8}$ for LPL), closely resembling the ${\theta }_{y}$ angular spectrum and peak value of electrons. The ${\theta }_{z}$ angular spectrum for different particles in the CPL pattern is similar to those of ${\theta }_{y}$ angular spectrum in the CPL pattern. Notably, for all particles in the LPL pattern, they are focused on the ${\theta }_{z}=$ 0 owing to the absence of electric field in the $Z$ direction. Particles in the CPL pattern display a significantly broader ${\theta }_{z}$ angular spectrum compared to those in the LPL pattern. The widths of the ${\theta }_{z}$ angular spectrum for particles are greater than that of the ${\theta }_{y}$ angular spectrum in the CPL pattern, with peaks of the ${\theta }_{z}$ angular spectrum approximately at $9.5\times 10^{10}$ for $\gamma$-ray beam, $1.2\times 10^{9}$ for electron beam, and $1.8\times 10^{8}$ for positron beam. These discrepancies could be attributed to limitations in the 2D simulation, which makes it difficulty to select the effective region in the $Z$ director.

% figure5

\subsection{\label{sec:parameter}Particle beams parameters}

The main parameters for the particle beams of simulated results, including the beam brilliance and emittance, are further discussed. The particle beam brilliance is given by Eq. (\ref{eq:L}),
\begin{equation}
L=\frac{N}{T\times D^2\times \theta^2 }\label{eq:L},
\end{equation}
where $N$ is the number of particles when the 2D area is extended to the 3D space, $T$ is approximately 30 fs, $D= 10~\mathrm{\mu m}$, and $\theta = a_0 / {\gamma }_{{e}^{-}}$ (Lorentz factor ${\gamma }_{{e}^{-}}\approx 1.57\times {10}^{4}$) is 12 mrad and 17 mrad for the CPL and LPL patterns, respectively \cite{sarri2014ultrahigh}.
The particle beam emittance in 2D space is described by,
\begin{equation}
\varepsilon = \sqrt{\left \langle y^{2}\right \rangle\left \langle \left ( \frac{p_{y}}{p_{x}}\right )^{2}\right \rangle - \left \langle y \tfrac{p_{y}}{p_{x}}\right \rangle^{2}}\label{eq:D}.
\end{equation}
The energy conversion efficiency is defined as $\eta = E_{p}/E_{le}$, where $E_{p}$ is the total energy of the particle and $E_{le}$ is the sum of the corresponding laser and initial electrons energies.

\begin{figure*}[htbp]
\includegraphics{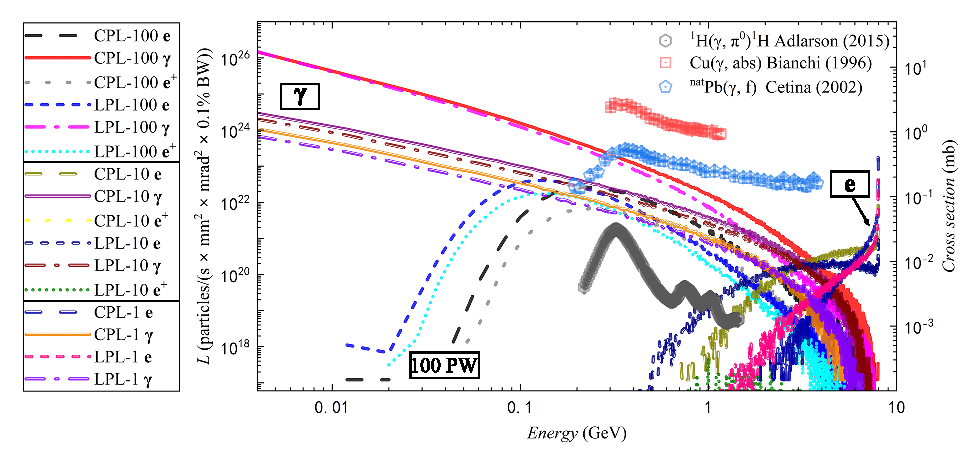}
\caption{(Color online) The particle beam brilliance spectrum (denoted by the left $Y$ axis) at 700 fs simulated by SMILEI for collision between the CPL and LPL patterns of SEL 100 PW, Shanghai Superintense Ultrafast Laser Facility (SULF) 10 PW and 1 PW lasers and the SLA 8 GeV electron beam, respectively. The  experimental data for photonuclear reaction cross-sections are denoted by the right $Y$ axis \cite{bianchi1996total, cetina2002photofission, adlarson2015measurement}.}
\label{L-pep}
\end{figure*}

For comparison, the 10 PW and 1 PW lasers at the SULF facility are also simulated, which collide on the same electron beam to yield particle beams. The 10 PW and 1 PW laser intensities are 10$^{21}$ W/cm$^{2}$ and 10$^{20}$ W/cm$^{2}$, correspondingly, with focal spot sizes of 5.5 $\mu$m and 40 $\mu$m, respectively, and FWHMs both of 30 fs \cite{zongxin20201, gan2021shanghai}. In Fig. \ref{L-pep}, the simulated particle beam brilliance spectrum and experimental cross-sections of particles \cite{bianchi1996total, cetina2002photofission, adlarson2015measurement} for the 1 PW and 10 PW lasers at SULF, as well as the 100 PW laser at SHINE on high-energy electron beam are plotted. The marvelous differences between the simulated results are that in the SULF 10 PW laser fewer positron can be produced, while in the SULF 1 PW laser cannot. One interesting finding for the SEL 100 PW laser is that the trends of the energy-brilliance for beam particles are similar to the trends of the energy spectra for all particles. Although in the different laser polarization states and the number of particles in the beam region, the peak energy of electrons and positrons in the CPL pattern is still higher than those in the LPL pattern. The maximum value of beam brilliance in the CPL pattern is slightly lower than that in the LPL pattern. In accordance with the beam brilliance of SEL 100 PW, SULF 10 PW and 1 PW lasers, the phenomena clearly demonstrate that more intense ultra-short laser, more $\gamma$-rays are produced by colliding with electron beam, and more positrons are also produced.

% figure6

% Table 1
\begin{table*}
\centering
\caption{\label{tab:ParaIonBeam}Particle beam parameters simulated by SMILEI for collision between circularly polarized (CPL) and linearly polarized (LPL) SEL 100 PW laser and SLA 8 GeV electron beam. The laser polarization (P), particle type (T), beam brilliance ($\mathit{L}$), average particle energy ($E_{avg}$), maximum particle energy ($E_{max}$), emittance ($\varepsilon$), and energy conversion efficiency ($\eta$) are given.}
\begin{ruledtabular}
\begin{tabular}{ccccccc}
P    & T     & $\mathit{L}$ (particles/(s mm$^{2}$ mrad$^{2}$)) & $E_{avg}$ (GeV) & $E_{max}$ (GeV) & $\varepsilon$ (mm mrad) & $\eta $ (\%)   \\ \hline
 & $\gamma$-ray    & $1.09\times 10^{27}$  & 0.232      & 7.950      & 0.109     & 4.47 \\
CPL & Electron & $1.52\times 10^{25}$  & 0.676      & 7.726      & 0.099     & 0.18 \\
 & Positron & $3.97\times 10^{24}$  & 0.633      & 6.598      & 0.101     & 0.05 \\ \hline
 & $\gamma$-ray    & $8.89\times 10^{26}$  & 0.158      & 7.975      & 0.165     & 5.96 \\
LPL & Electron & $1.05\times 10^{25}$  & 0.418      & 7.727      & 0.156     & 0.19 \\
 & Positron & $4.93\times 10^{24}$  & 0.421      & 6.860      & 0.152     & 0.09 
\end{tabular}
\end{ruledtabular}
\end{table*}

Table~\ref{tab:ParaIonBeam} shows the parameters of particle beams simulated by SMILEI for collisions between SEL 100 PW and SLA 8 GeV electron beam according to Fig.~\ref{L-pep}. The $\gamma$-ray beam brilliance in the CPL pattern (up to 1.09 $\times$ 10$^{27}$ photons/(s mm$^{2}$ mrad$^{2}$)) is higher than that in the LPL pattern, and the positron beam brilliance in the LPL pattern (up to 4.93 $\times$ 10$^{24}$ positrons/(s mm$^{2}$ mrad$^{2}$)) is higher than that in the CPL pattern. Interestingly, the beam emittance is around 0.1 mm mrad in the CPL pattern, which is much smaller than 0.2 mm mrad in the LPL pattern.

In the CPL pattern, the $\gamma$-rays are most abundant in the beam region, followed by electrons, and positrons are the least numerous, with electron beam brilliance being 1.52 $\times {10}^{25}$ electrons/(s mm$^{2}$ mrad$^{2}$) and positron beam brilliance being 3.97 $\times {10}^{24}$ positrons/(s mm$^{2}$ mrad$^{2}$). Similarly, the $\gamma$-ray energy conversion efficiency of 4.47$\%$ is the maximum, and the energy conversion efficiencies of electrons and positrons are 0.18$\%$ and 0.05$\%$, respectively. The $\gamma$-rays and electrons have the highest energy in beam region (up to 8 GeV), while the maximum particle energy of positron beams is approximately 7 GeV. The average particle energy of electron and positron beams, which is around 0.6 GeV, is higher than that of $\gamma$-ray beam ($\approx$ 0.2 GeV).

In the LPL pattern, the beam brilliance, energy conversion efficiency, maximum particle energy, and average particle energy of $\gamma$-rays, electrons and positrons have the same numerical order. The $\gamma$-ray and electron beam brilliants are 8.89 $\times {10}^{26}$ photons/(s mm$^{2}$ mrad$^{2}$) and 1.05 $\times {10}^{25}$ electrons/(s mm$^{2}$ mrad$^{2}$). The energy conversion efficiencies of $\gamma$-rays is 5.96$\%$, which is marginally superior to 2$\%$ given by the all-optical scheme for laser energy into $\gamma$-rays in 2012 \cite{thomas2012strong}. The energy conversion efficiencies of electrons and positrons are 0.19$\%$ and 0.09$\%$, respectively. The maximum particle energies are similar to those in the CPL pattern. The average particle energy of electron and positron beams is about 0.4 GeV, and average particle energy of $\gamma$-ray beam is about 0.2 GeV. Although the maximum energy of $\gamma$-rays approaches the initial electron energy during the collision process, the majority of $\gamma$-rays still have lower energy, while most of the positrons have higher energy.

In general, the energy conversion efficiency of particles in the CPL pattern is lower than that in the LPL pattern, the maximum particle energy in the CPL pattern is nearly equal to that in the LPL pattern, and average particle energy in the CPL pattern is higher than that in the LPL pattern. The conflicting results could be related to the fact that more positrons are generated, with lower energy and greater spatial emittance in the beam region in the LPL pattern. It is suggested to generate $\gamma$-rays and electrons demonstrate superior beam quality in the CPL pattern, whereas to generate positrons higher beam quality in the LPL pattern at the SHINE facility.

\subsection{\label{sec:device}Facility layout and applications}

The simulations in this work provide an alternative way to produce high-quality $\gamma$-ray and positron beams of the all-optical scheme designation \cite{zhu2016dense, lobet2017generation, zhu2018generation}, which uses the UIUSL to generate a higher-energy electron beam which then collide with another UIUSL. The simulated results are aimed at assessing the feasibility and advantages of electron-laser colliding $\gamma$-ray source and positron beam facility, which provides new opportunity to the high-energy photonuclear reactions and positron related physics within the aforementioned energy ranges. Examples of different ($\gamma, \pi^{0}$), ($\gamma$, abs) and ($\gamma,$ f) reactions within the high-energy scopes are also plotted in Fig.~\ref{L-pep}, which could enhance the comprehensive understanding of nuclear structures such as a form factor, polarization, and resonances \cite{ahrens2001half}, and the high-brilliance and high-flux of the particle beams may partly advance the nuclear reaction yield. Particle energies follow a continuous spectrum defined by the surrounding temperature in the high-temperature, high-density plasma of extreme astrophysical environments, including AGN/GRB jets, pulsars, stellar interiors, supernovae, neutron star mergers, and so on. Utilizing a beam with a continuous energy spectrum could more effectively simulate the overall effects of particles in real astrophysical environments, such as the integrated reaction rates under a continuous photon spectrum. The potential high-quality $\gamma$-ray beams at the SHINE facility would significantly promote the nuclear reaction researches, as well as strange phenomena in nuclear physics and nuclear astrophysics.

% figure7
\begin{figure}[htbp]
\includegraphics{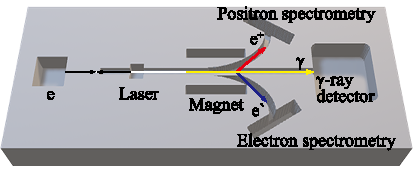}
\caption{(Color online) A Sketch drawing of a colliding station and its high-energy $\gamma$-ray beamline and high-energy positron beamline based on the SEL 100 PW laser and SLA 8 GeV electron beam at the SHINE facility.}
\label{layout}
\end{figure}

The concept design for a possible colliding station based on the SEL 100 PW laser and SLA 8 GeV electron beam at the SHINE facility is drawn in Fig.~\ref{layout}. As demonstrated by the simulated results, the spatial distributions of three types of particles overlaps. Therefore, it is possible to achieve successful beam separation by placing deflecting magnets above and below the beam region \cite{abramowicz2021conceptual, crippa2023quantum, Brun2024}. Neutral $\gamma$-rays would still propagate in the positive direction along the $X$ axis, while electrons and positrons would propagate in opposite directions along the $Y$ axis \cite{burke1997positron, zhu2016dense, yu2019creation}. The applications of $\gamma$-rays in different energy regions are as follows,
\begin{itemize}
\item High-energy $\gamma$-rays ($\geqslant 140~\mathrm{MeV}$) to research pion photoproduction on the nucleon \cite{enoch1957theory, zhao2002pion, tarbert2014neutron}, and determine more accurate resonance coupling constants through more experimental measurements of the nucleon resonance spectrum \cite{klempt2010baryon}.
\item Medium-energy $\gamma$-rays ($\geqslant$ 100 MeV) to study photon nuclear spallation reactions etc. for improving the photonuclear reaction database \cite{shibata1987photonuclear, haba2003systematic, wei2023spagins}.
\item Low-energy $\gamma$-rays (1-30 MeV) to investigate the laboratory astrophysics \cite{ong2019feasibility, takabe2021recent}, and the enhanced low-energy $\gamma$-rays also make it strong medical isotopes production factory \cite{pang2023MedisotopeSLEGS, gu2024feasibility} and photo-transmutation \cite{magill2003laser, jin2008transmutation, wang2016photo} through photonuclear reactions \cite{weller2009research, beard2012photonuclear, nedorezov2021nuclear}.
\end{itemize}

In concrete terms, the above high-quality particle beams may have practical multidisciplinary applications of
\begin{itemize}
\item Photonuclear reactions and related applications. The different cross-sections of photonuclear reactions induced by high-energy $\gamma$-rays allow for the differentiation between stable and unstable isotopes in samples using $\gamma$-ray spectroscopy \cite{budker2022expanding, liu2022novel, yang2023novel}, radioactive isotopes productions for medical applications \cite{pang2023MedisotopeSLEGS, LiZC2023NST, YYX24RadPhChem}, and reaction mechanism of nuclear spallation reactions \cite{wei2023spagins}. These researches are in potential be performed within $\gamma$-ray energy ranging from hundreds MeV to 8 GeV. Meanwhile, the strong penetrating power of $\gamma$-rays can be applied for non-destructive testing of large engineering samples, while their high energy can serve as a microscope for the study of material lattice dynamics \cite{hajima2014application, zhou2023optimization, li2024two}.
\item Positrons and related applications. As one type of antimatter particles, the positrons have applications in the research of particle physics \cite{apyan2023gamma}, solid-state physics, such as diagnosing complex structures of Fermi surfaces \cite{thirolf2014bright}, and Positron Emission Tomography (PET) for material sciences \cite{ledingham2004high}.
\item Nonlinear strong-field QED verification. Experiments for precision measurements of high-energy heavy element reaction events can be conducted to verify the predictions of nonlinear QED effects using these three types of particles \cite{elkina2011qed, mackenroth2019nonlinear, adam2021measurement}.
\end{itemize}

Several questions still remain in techniques. One is how to achieve collisions by synchronizing the electron beam and laser within a few femtoseconds and microns \cite{galynskii2001nonlinear, kharin2018higher, zhang2019quantum}, which requires extremely high initial beam quality and technical precision \cite{seipt2011scaling, albert2016applications, krafft2023scattered}. The other one is that the maximum energy of the three types of particles may not exceed the maximum energy of the initial electron beam, which requires to increase the electron beam energy of the accelerator to obtain higher $\gamma$-ray and positron beam energies.

\section{\label{sec:conclusion}Conclusion}

A comprehensive simulation has been performed on the particle productions in collisions between the SEL 100 PW laser and SLA 8 GeV electron beam based on the SHINE facility by the PIC program SMILEI. Most of the $\gamma$-rays and electron-positron pairs generated by NICS and MBWP are found to be localized in the beam center, where the beam center size matches that of the initial electron pulse. The majority of $\gamma$-rays are distributed in the lower energy range ($Energy <$ 0.9 GeV), and the generated electrons and positrons are concentrated at $Energy = $ 0.2 GeV for CPL and $Energy = $ 0.1 GeV for LPL. Electrons and positrons move forward in the beam center, whereas those in the ``tail end" spread out in 3D space for the CPL pattern and spread out in the $X$ and $Y$ directions for the LPL pattern. Within the selected beam region, the angles of the particle momentum are clustered around 0 degrees, and the beam brilliance rises with increasing laser intensity. The beam brilliance of the $\gamma$-rays, with a maximum energy of up to 8 GeV, reaches 10$^{27}$ photons/(s mm$^{2}$ mrad$^{2}$). The beam emittance is approximately 0.1 mm mrad in the CPL pattern, and its energy conversion efficiency is up to 5.96$\%$ in the LPL pattern. The maximum energy of positrons reaches 7 GeV, with a beam brilliance of up to 4 $\times$ 10$^{24}$ positrons/(s mm$^{2}$ mrad$^{2}$), a beam emittance of approximately 0.1 mm mrad in the CPL pattern, and a energy conversion efficiency of 0.09$\%$ in the LPL pattern. The high brilliance, high energy, and low emittance are the advantages of the $\gamma$-ray, positron and electron beams, which are generated in the simulated collisions. The $\gamma$-ray beam brilliance is higher and the beam emittance is lower in the CPL pattern than in the LPL pattern, and the positron beam brilliance is higher and the particle energy conversion efficiency is larger in the LPL pattern than in the CPL pattern. The insights gained from these results may inspire us to choose distinct laser polarization patterns, various energy ranges achieved through moderating materials, and diverse particle beams guided by magnets to achieve specific goals of experimental research, such as studying photonuclear reactions and verifying strong-field QED theory.

Besides the quantitative and comprehensive assessment of top-quality $\gamma$-ray and positron beams with the most intense laser, this work explores the feasibility of the facility design scheme. Multiple frontier applications have been proposed with these $\gamma-$ray and positron beams, including $\gamma$-ray microscopy in material science and engineering, $\gamma$-ray spectrum diagnostics of radionuclides, PET for advanced nuclear analysis technology, and so on. Moreover, if the particle beams could be polarized \cite{potylitsyn2017photon, king2020nonlinear, liu2023simulation} and vortexed \cite{lu2023manipulation}, they have significant potential to enhance the research dimension of plasma physics, and advance studies in nuclear physics and astrophysics.

\section*{Data Availability}
The data that support the findings of this study are available from the corresponding authors upon reasonable request.

\begin{acknowledgments}
We are indebted to Prof. Wenqing Shen for his significant proposal and discussion on this work. This work is supported by the Strategic Priority Research Program of the CAS under Grant No. XDB34030000, the National Key Research and Development Program of China (No. 2022YFA1602404, No. 2022YFA1602402), the National Natural Science Foundation of China (No. 12475134, No. 12235003, No. U1832129), the Youth Innovation Promotion Association CAS (No. 2017309), and Natural Science Foundation of Henan Province (Grant No. 242300421048).
\end{acknowledgments}

\appendix

\end{document}